\documentclass[notitlepage, twocolumn]{revtex4-1}

\usepackage{verbatim}
\usepackage{amsmath}
\usepackage{amsfonts}
\usepackage{amssymb}
\usepackage{amsthm}
\usepackage{bm}
\usepackage{xparse}
\usepackage{graphicx}
\usepackage{xcolor}
\usepackage{textcase}
\usepackage{url}
\usepackage{lipsum}
\usepackage{appendix}
\usepackage{epstopdf}
\usepackage{parskip}

\newcommand{\avg}[1]{\langle{#1}\rangle} 
\newcommand{\ket}[1]{| {#1} \rangle} 
\newcommand{\bra}[1]{\langle {#1} |} 
\DeclareDocumentCommand{\Tr}{m m O{\big}}{{\rm Tr}_{\:\!{#1}}#3({#2}#3)}

\DeclareMathOperator*{\tr}{\mathrm{Tr}}

\newtheorem{theorem}{Theorem}

\begin{document}
\title{Single-copy entanglement detection}
\author{Aleksandra Dimi\'{c}}
\affiliation{
Faculty of Physics,
University of Belgrade,
Studentski Trg 12-16,
11000 Belgrade, Serbia
}
\author{Borivoje Daki\'c}
\affiliation{
Institute for Quantum Optics and Quantum Information (IQOQI),
Austrian Academy of Sciences, Boltzmanngasse 3,
A-1090 Vienna, Austria
}
\affiliation{
Vienna Center for Quantum Science and Technology (VCQ), Faculty of Physics, Boltzmanngasse 5, University of Vienna, Vienna A-1090, Austria
}

\date{\today}

\begin{abstract}
One of the main challenges of quantum information is the reliable verification of quantum entanglement. The conventional detection schemes require repeated measurement on a large number of identically prepared systems. This is hard to achieve in practice when dealing with large-scale entangled quantum systems. In this letter we formulate verification as a decision procedure, i.e. entanglement is seen as the ability of quantum system to answer certain ``yes-no questions''. We show that for a variety of large quantum states even a single copy suffices to detect entanglement with a high probability by using local measurements. For example, a single copy of a $16$-qubit $k$-producible state or one copy of $24$-qubit linear cluster state suffices to verify entanglement with more than $95\%$ confidence. Our method is applicable to many important classes of states, such as cluster states or ground states of local Hamiltonians in general.
\end{abstract}

\maketitle

\section*{INTRODUCTION}
A main focus of modern practical quantum information research is on the generation of large-scale quantum entanglement involving many particles with the goal of achieving real applications of quantum technologies~\cite{Castelvecchi,Dow03}. Recent quantum experiments dealing with a large number of particles, such as optical lattice simulations involving $10^3-10^4$ atoms~\cite{Yan13,Hart15,Takei16,Campbell17}, experiments with hundreds of trapped ions~\cite{Britton12} or thousands of qubits in D-Wave systems~\footnote{See: https://www.dwavesys.com/}, show the real potential for applications of quantum technologies in the near future. An important instance of this challenge is the verification problem, i.e. how to reliably detect the presence of quantum resources, in particular quantum entanglement. The plausibility of standard verification schemes is questionable, since they require repeated measurement on large ensemble of identically prepared copies, which is highly demanding to achieve in practice.

One way of detecting quantum entanglement is to perform full quantum state tomography~\cite{Bisio09} (see also \cite{Xin16,Carmeli16}) from which one can extract full information about the quantum state preparation (one can recover the entire density matrix). However, the full tomography becomes an unrealistic task already for a moderate size of quantum systems as the number of required measurement settings grows exponentially fast with the size of system. Luckily, in many cases, the knowledge of the entire quantum state is not needed, i.e. one can witness the presence of entanglement by measuring the mean values or  the higher moments of a moderate number of physical quantities (observables). This is the groundwork for detection methods based on witness operators~\cite{Hor96,Ter03,Brukner04,Toth05,Dowling04,Knips16}, non-linear entanglement witnesses~\cite{Guh07,Pat08,Arrazola12}, Bell's inequalities~\cite{Navascues16,Acin16}, quantum Fisher information~\cite{Smerzi09,Hyllus12,Toth12,Smerzi16} and random correlations~\cite{Tran15,tran16,Bruss15}. These methods have been proven extremely useful for many practical situations and they have been extensively developed for the detection of a variety of quantum states and adapted to various scenarios (see review articles ~\cite{Guh09,Hor09,Sar14}). Nevertheless, all of the existing detection schemes are based on an idealized situation which requires \emph{repeated measurements on a large ensemble of independent and identically distributed (i.i.d.) copies of a quantum state}.

The typical detection procedure involves the extraction of the mean value of a certain witness operator $W=\sum_iW_i$, by measuring the means of the local observables $W_i=A_1\otimes\dots\otimes A_N$ ($N$ is the size of system).
Therefore, in order to detect entanglement, one has to conduct different experiments (different measurement settings for each $W_i$), each of which requires a large number of identically prepared copies such that the sample averages are close to the real mean values. However, as any practical situation deals with a finite amount of data generated by a quantum measurement, the presence of entanglement can be verified only with a certain level of confidence, thus an adequate statistical analysis is necessary~\cite{Jungnitsch10,Kohout10,Arrazola13}. Moreover, it is very hard to fulfil these requirements when dealing with large-scale entangled systems as only a limited (rather low) number of instances of a given quantum resource is available, due to various technical challenges, such as the lack of a good control and manipulation. As an example, one can take a recent experiment done with single photons~\cite{Wang15} where a 10-photon coincidence was registered every five minutes in average. With the same technology, that is by using the parametric down-conversion and postselection techniques, every additional photon pair (e.g. 12, 14, $\dots$ photons) would reduce the count rate by at least one order of magnitude, consequently making the duration of the experiment (to verify entanglement) months or even years longer. In such situations, where only a limited number of resources is available, the main question arises whether it is still possible to reliably and efficiently certify the presence of quantum entanglement? This question is not only interesting from the theoretical point of view, it is also of great importance for practical quantum information.

In recent years we have seen several works that go beyond i.i.d. scenario, in the context of quantum state tomography~\cite{Renner11} and reliable entanglement verification~\cite{Arrazola12,Arrazola13}. Although the techniques and methods developed there are quiet generic, they still require a large sample sizes to verify entanglement with high confidence. On the other hand, in situations where only a low number of instances of a given quantum resource are available it appears natural to employ \emph{random sampling techniques}~\cite{Till06} for reliable detection. The advantage of such methods comes
from simplicity of the data analysis, since minimal prior knowledge of the global population is needed. In the quantum scenario, random sampling has been proven very useful for quantum communication complexity~\cite{Buh10,Sca14}, tomography via compressed sensing~\cite{Eis10}, fidelity estimation~\cite{Fla11}, self-testing methods~\cite{Yao03,McKague10,Sca12,Shi12,Vaz12,McKague12}, quantum state certification~\cite{Eis15,Eis17}, quantum secret sharing~\cite{Mark14} and verification of quantum computing~\cite{Fitz17}. Some of these methods can be used to verify entanglement probabilistically as demonstrated in \cite{Ior12,Tame16}. In the same spirit, we shall incorporate random sampling methods together with techniques of quantum communication complexity~\cite{Buh10,Sca14} to propose entanglement verification scheme in the form of \emph{a quantum information task}. Unlike focusing on a large ensemble of i.i.d. copies, our main target here is a single experimental run, i.e. the central quantity for entanglement detection is \emph{the probability of success} to perform certain binary task, given that the state was entangled/separable. Therefore, our scheme is designed to detect entanglement probabilistically. This framework has two main advantages as compared to conventional detection schemes:
\begin{itemize}
\item[a)] \emph{it promises a dramatic reduction of the resources needed for reliable verification in large quantum systems}, and
\item[b)] \emph{it provides a simple tool for reliable statistical analysis of errors and confidence intervals}.
\end{itemize}
Most importantly, we show that in many situations the probability of success (to accomplish certain binary task) decreases exponentially fast with the size of system for all separable input states, whereas it approaches certainty if a particular entangled state (the target state) was prepared. Thus, even a \emph{single experimental run (single copy) can reveal the presence of entanglement with high accuracy}. To our best knowledge, this is the first demonstration of entanglement detection in a single-copy regime (apart from the well-studied example of an i.i.d. state $\rho^{\otimes N}$, see example of $k$-producible state bellow). We explicitly construct the detection procedure for $k$-producible states~\cite{Brig05} and cluster states~\cite{Raussendorf01}. We show that, for example, one copy of a $16$-qubit $k$-producible state or a single-copy of $24$-qubit linear cluster state suffices to certify entanglement with more than $95\%$ confidence. Thus, our method is applicable for quantum experiments involving tens of entangled qubits with the promise of a dramatic reduction of the resources needed for reliable entanglement detection (as compared to standard methods). Furthermore, the method developed for $k$-producible states can be used to naturally embed the standard techniques based on entanglement witnesses into our framework, meaning the statistical analysis of confidence intervals and errors simple and straightforward. Finally, we develop a general method for entanglement detection in ground states of local Hamiltonians (that exhibit the so-called entanglement gap~\cite{Dowling04}). These include many important classes of quantum states, such as the matrix product states~\cite{Verstraete07} and projected-entangled pair states~\cite{Verstraete08} as they can be seen as unique ground states of the so-called parent Hamiltonians~\cite{Verstraete07,Perez-Garcia07}. At the end, we analyze the noise effect and we show that our probabilistic detection is very robust against the noise modeled by an arbitrary separable state.

\begin{figure}[]
\centering
\includegraphics[width=8cm]{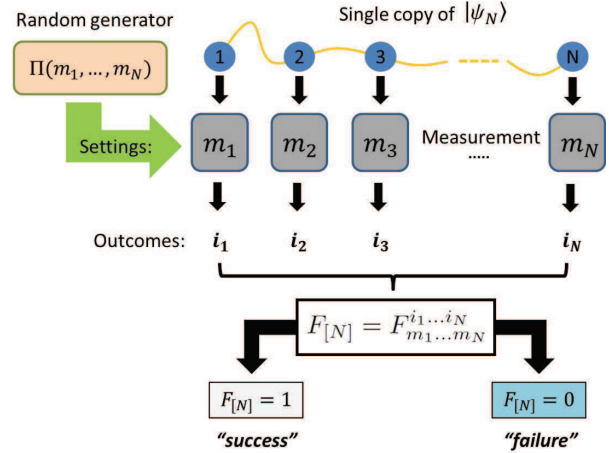}
\caption{{\bf Probabilistic entanglement detection}. A single-copy of $N$-partite quantum state is prepared. The sequence of measurement settings $\{m_1,\dots,m_N\}$ is randomly drawn from distribution $\Pi(m_1,\dots,m_N)$. Each $m_k$ is locally executed on $k$th subsystem and the set of outcomes $\{i_1,\dots,i_N\}$ is obtained. The value of binary cost function $F_{[N]}=F_{m_1\dots m_N}^{i_1\dots i_N}$ prescribes either ``success'' ($F_{[N]}=1$) or ``failure'' ($F_{[N]}=0$) to the experimental run.}
\label{fig:1}
\end{figure}

\section*{RESULTS}
\subsection*{Detection framework}
Our method relies on a decision procedure where entanglement is seen as the ability of quantum systems to answer certain ``yes-no questions''. The main figure of merit is the probability of success that a certain binary cost function $F$ evaluates to 1, i.e. $P[F=1]$. The main goal is to provide examples of quantum states where $P$ decreases exponentially fast to zero as the size of system grows for all separable states, whereas it approaches certainty ($P=1$) if a particular entangled state (the target state) was prepared. Therefore, one can verify the presence of quantum entanglement with high probability even by measuring a single copy of a large quantum system.

In order to explain how our scheme works we consider a quantum system composed of $N$ subsystems each residing in a finite-dimensional Hilbert space of dimension $d$. We usually assume $N$ is large, although all the formulas derived hold for general $N$. To each subsystem we associate a certain set of possible measurements that can be performed locally. For example, in the case of qubits we may chose to measure each of them in complementary bases (such as $X$ and $Z$ measurement). In the general case, we shall include the most general quantum measurements (POVMs). Thus to each subsystem we associate a set of $M$ different measurement settings defined by the set of positive semidefinite operators $E^{(k)}_{mi}$, where $\sum_{i}E^{(k)}_{mi}=\openone$ and $m=1\dots M$. Here $k$ labels the subsystem, $m$ the measurement setting and $i$ labels the measurement outcome.

Given a single-copy of an $N$-partite quantum system, the detection procedure consists of the following four steps (see Figure 1):
\begin{itemize}
  \item[1.] A sequence of measurement settings $\{m_1,m_2,\dots,m_N\}$ is randomly generated from the probability distribution of settings $\Pi(m_1,\dots,m_M)$,
  \item[2.] The measurements are locally executed on each subsystem and the set of outcomes $\{i_1,\dots,i_N\}$ is obtained,
  \item[3.] A certain binary ($0/1$) cost function of settings and outcomes $F_{[N]}=F^{i_1\dots i_N}_{m_1\dots m_N}$ is computed,
  \item[4.] If $F_{[N]}=0/1$ we associate ``success/failure'' to the experimental run.
\end{itemize}
This is the way to establish the probabilistic framework for entanglement detection trough the probability of success $P[F_{[N]}=1]$. Our goal here is to choose the cost functions such that the probability of success vanishes exponentially fast in $N$ for all separable states $\rho_{sep}$
\begin{equation}\label{sep bound}
P_{\rho_{sep}}[F_{[N]}=1]\leq\exp[-Nc],
\end{equation}
where $c>0$ is some constant. On the other hand, the cost function $F_{[N]}$ is chosen such that there is an entangled state for which $P_{\rho_{ent}}[F_{[N]}=1]\approx1$, meaning that whenever the state $\rho_{ent}$ (target state) has been prepared, the detection scheme works even in a single-copy scenario. An explicit bound on the probability of success for concrete examples will be derived.

\subsection*{Example of $k$-producible state}
A good example to start with is that of the $k$-producible entangled state~\cite{Brig05}, i.e. $\ket{\phi_1}\ket{\phi_2}\dots\ket{\phi_m}$, where the products $\ket{\phi_s}$ involve at most $k$ parties. For simplicity, we chose the target state to be the product of quantum singlets $\ket{\psi_0}=\ket{\psi^{-}}^{\otimes N}$, where $\ket{\psi^{-}}=\frac{1}{\sqrt2}(\ket{01}-\ket{10})$~\footnote{The example and task presented here is quiet similar to ``non-local'' quantum games (see \cite{Buh10,Sca14}). Nevertheless, we believe it is a good starting point for introducing a more delicate examples that follow in subsequent sections.}. We consider the set of $\{X,Y,Z\}$ measurement settings for each qubit, meaning that the measurement is performed in the eigenbasis of Pauli operators with the set of binary outcomes $i=0,1$. The quantum singlet is the unique state for which $X\otimes X=Y\otimes Y=Z\otimes Z=-1$, meaning that the measurement of $X\otimes X$, $Y\otimes Y$, and $Z\otimes Z$ reveals perfect anticorrelations. Let us introduce the projectors on the outcome $-1$ for the correlation measurement
\begin{equation} \label{singlet projector X}
Q=\frac{\openone-X\otimes X}{2}
\end{equation}
\begin{equation} \label{singlet projector Y}
W=\frac{\openone-Y\otimes Y}{2}
\end{equation}
\begin{equation} \label{singlet projector Z}
R=\frac{\openone-Z\otimes Z}{2}.
\end{equation}
To these projectors we associate three measurement settings $S=\{XX,YY,ZZ\}$. Although, the projectors are commutative, there is no separable state for which the measurement reveals $Q=W=R=1$ simultaneously (this is a property of the singlet state only). Therefore, if we pick one of the settings from $S$ randomly (with probability $1/3$), there is a chance of at most $2/3$ to get the outcome $1$ for all separable inputs. More precisely, in such a case, the probability of success
\begin{equation}\label{singlet Fmean}
P_{\rho_{sep}}=\avg{\frac{1}{3}(Q+W+R)}\leq\frac{2}{3},
\end{equation}
for all separable two-qubit states $\rho_{sep}$. Here $\avg{\cdot}=\tr(\cdot)\rho$ denotes the mean value. This observation clearly suggests a detection scheme. We shall divide the set of $2N$ qubits into consecutive pairs and for every individual pair we pick one of the settings from $S$ randomly (with probability $1/3$) and perform the corresponding correlation measurement \eqref{singlet projector X}-\eqref{singlet projector Z}. For separable inputs, the bound \eqref{singlet Fmean} suggests that the relative frequency of the outcome $1$ cannot significantly exceed the value of $2/3$ (provided that $N$ is large). Formally, we define the frequency $R_{[N]}=\sum_{k=1}^{N}F_k$, where $F_k$ is the outcome of the correlation measurement on individual pairs
\begin{equation}
F_k=\frac{1}{2}\left(1-(-1)^{i_{k}+j_{k}}\right).
\end{equation}
Here $i_k,j_k=0,1$ label the single-qubit measurement outcomes for the $k$th pair. The cost function is defined as
\begin{equation}
F_{[N]}=\left\{
          \begin{array}{ll}
            1, & \hbox{$R_{[N]}\geq (\frac{2}{3}+\delta)N$;} \\
            0, & \hbox{$R_{[N]}< (\frac{2}{3}+\delta)N$,}
          \end{array}
        \right.
\end{equation}
where $\delta>0$ is some constant we keep at the moment as a free parameter. In other words, we associate ``success'' to the run if the number of local successes $F_k$ exceeds certain threshold of $(\frac{2}{3}+\delta)N$. The overall probability of success reads
\begin{equation}\label{singlet Ps}
P_{\rho}[F_{[N]}=1]=P_{\rho}\left[F_{1}+\dots+F_N\geq \left(\frac{2}{3}+\delta\right) N\right],
\end{equation}
and we recognize in the last equation the probability that the sum of random variables $F_1+\dots+F_N$ exceeds the value of $(\frac{2}{3}+\delta) N$. If the input state is a product state $\rho_{prod}=\rho_1\otimes\dots\otimes\rho_{2N}$, the random variables $F_k$ are independent with $\avg{F_k}\leq\frac{2}{3}$. For such a case the bound on \eqref{singlet Ps} is well studied in classical probability theory and the results are known as the Chernoff bounds~\cite{Chernoff}. We show in the Appendix that
\begin{equation}\label{singlet sep bound}
P_{\rho_{prod}}[F_{[N]}=1]\leq e^{-D(\frac{2}{3}+\delta||\frac{2}{3})N},
\end{equation}
where $D(x||y)=x\log{\frac{x}{y}}+(1-x)\log{\frac{1-x}{1-y}}\geq0$ is the Kullback--Leibler divergence. Furthermore, if the bound holds for all product states, it also holds for their mixtures, i.e. it holds for all separable states. We see that the probability of success vanishes exponentially fast in $N$, for all $\delta>0$. This is quite convenient, as we do not have to fix $\delta$ in advance. Once the experiment has been performed, we can calculate directly from the experimental data $F_1,F_2,\dots F_N$ how much the frequency deviates from $2N/3$, i.e. we set $\delta=(F_1+\dots+F_N)/N-2/3$, and consequently calculate the bound on probability of success for separable states by using \eqref{singlet sep bound}.
For the case of $\ket{\psi_0}=\ket{\psi^{-}}^{\otimes N}$ input state, each local cost function $F_k=1$ deterministically, thus we get $\delta=1/3$. The bound \eqref{singlet sep bound} reduces to
\begin{equation}
P_{\rho_{sep}}[F_{[N]}=1]\leq \left(\frac{2}{3}\right)^N.
\end{equation}
If $N$ is sufficiently large,  a single-copy of $\ket{\psi_0}$ suffices to certify presence of entanglement with high probability. For example, if we want to have a detection probability of at least $95\%$ (i.e. we want to be sure that no separable state has a probability of success more than 5\%) in a single-shot experiment, we get the minimal number $N_{min}=8$, which is a remarkably low number.

The present example shows how the standard detection of entanglement in quantum singlet based on the witness operator $\frac{1}{3}(Q+W+R)$ can be naturally embedded in our framework. Conventionally, one has to measure $Q$, $W$ and $R$ in three separate experiments by using an i.i.d. ensemble of qubit pairs $\rho_{12}^{\otimes N}$ in order to estimate the mean values $\avg{Q}$, $\avg{W}$ and $\avg{R}$. However, the i.i.d. assumption is difficult to justify operationally, hence the statistical analysis involving many experiments is non-trivial. Furthermore, if the number of singlet pairs is low, it not clear how to actually pursue the detection scheme.  For example, imagine a situation where only $N=8$ pairs are available. The question is how to divide the pairs and perform the corresponding measurements. We may use the first three copies to measure $Q$, the second three to measure $W$, and the last two for the measurement of $R$. However, if the order is known and fixed in advance, than the following product state $(\ket{x+}\ket{x-})^{\otimes3}(\ket{y+}\ket{y-})^{\otimes3}(\ket{z+}\ket{z-})^{\otimes2}$ gives exactly the same result as the i.i.d. state $\ket{\psi^{-}}^{\otimes 8}$. Thus, we cannot conclude the presence of entanglement or we may even wrongly claim its presence. Certainly, a correct statement requires a proper statistical analysis. On the other hand, one of the key procedures in our method is the random sampling of measurement settings, which provides us a simple tool to analyze the errors and confidence intervals trough the probability of success. Therefore, there is a clear separation between the state $\ket{\psi^{-}}^{\otimes 8}$ and the product state given above, as the later has only the chance of $(2/3)^8\approx0.039$ to reveal the result $F_{1}+\dots+F_8=8$.

In general, any detection based on witness operator can be incorporated in our framework, with the goal to achieve more resource-efficient entanglement detection. For a witness $W=\sum_iW_i$, one has to sample the measurements of $W_i$'s randomly every single experimental run. The bound similar to \eqref{singlet sep bound} can be easily derived. Nevertheless, one may object that in such a case, one still requires many copies for reliable detection (i.e. $N$ copies of $k$-partite state $\ket{\psi}$ folded into a single multipartite copy $\ket{\psi}^{\otimes N}$ ). In the next examples we will unambiguously show that, indeed, one can certify entanglement by measuring only a single-copy.

\subsection*{Example of cluster states}
Another example we present here is that of cluster states~\cite{Raussendorf01}. In contrast to the previous example of $k$-producible states, cluster states contain genuine multipartite entanglement~\cite{Eisert04} and they are known to be a universal resource for measurement-based quantum computation~\cite{MBC}. For the sake of simplicity, we shall explain how the single-copy detection scheme works for the case of linear cluster states (LCS). The generalization to higher dimensions is straightforward and we briefly discuss it at the end of this section. The $N$-qubit LCS is uniquely defined by the set of $2^N$ stabilizers, i.e.
\begin{equation}\label{stab equation}
G_{q_1\dots q_N}\ket{LCS}=G_1^{q_1}\dots G_N^{q_N}\ket{LCS}=+1\ket{LCS},
\end{equation}
where $G_k=Z_{k-1}X_kZ_{k+1}$ and $q_k=0,1$. Here $\{X_k,Y_k,Z_k\}$ represent the set of standard Pauli matrices acting on $k$th qubit and, for simplicity, we have chosen the cluster state with periodic boundaries, i.e. $Z_{N+1}\stackrel{\mathrm{def}}{=}Z_1$ and $X_{N+1}\stackrel{\mathrm{def}}{=}X_1$. We consider the set of $\{X,Y,Z\}$ measurement settings for each qubit, meaning that the measurement is performed in the eigenbasis of Pauli operators with the set of binary outcomes $i=0,1$. We start with a simple analysis by considering a small subset (cluster) of four qubits, let say $\{1,2,3,4\}$ with the corresponding stabilizers $G_2=Z_1X_2Z_3$, $G_3=Z_2X_3Z_4$ and $G_2G_3=Z_1Y_2Y_3Z_4$ acting solely on it. Although, all three stabilizers are commutative, they are not locally compatible (in a sense that there is no local measurement of all three of them simultaneously), therefore there is no product (separable) state for which all three observables can take the same value, i.e. $G_2=G_3=G_2G_3=+1$, simultaneously. For that reason, if we randomly chose (with probability $1/3$) to measure one of the stabilizers there is only a chance of $2/3$ to get the result $+1$, for all separable inputs (similar to the previous example of singlet state). This is the key observation that enables our detection method to work. Our main idea is to show that if we pick a random partition of the set of $N$ qubits into $4$-qubit clusters and subsequently on each of them randomly measure one of the corresponding stabilizers,  the relative frequency of the outcome $+1$ cannot significantly exceed the value of $2/3$. More formally, we start by introducing partitions of $N$ qubits into $4$-qubit clusters $\{c_{t_1},c_{t_2},\dots c_{t_L}\}$, where $c_{t_s}$ is the cluster involving the sequence of four qubits $c_{t_s}=\{t_s,t_s+1,t_s+2,t_s+3\}$. Furthermore, we see from the previous analysis that the border qubits in each cluster are always measured in the $Z$-basis (when measuring the corresponding stabilizer). Thus, we shall allow for possible overlap between neighbouring clusters on border qubits. More precisely, we say that the partition is regular if the neighbouring clusters overlap on at most one (border) qubit, i.e. $t_{s+1}-t_{s}\geq3$. For example, the partition $\{\dots,\{7,8,9,10\},\{10,11,12,13\},\dots\}$ is considered regular, whereas $\{\dots,\{7,8,9,10\},\{9,10,11,12\},\dots\}$ is irregular, as the two clusters in partition overlap on qubits 9 and 10 (see SI for more examples). We denote the set of all regular partitions of size $L$ by $\mathcal{C}_L$. We shall think of $L$ as being large, e.g. on the same scale as $O(N)$ with the number of qubits and, at the same time, we shall choose $L$ such that the set $\mathcal{C}_L$ is large in size as well. The clusters in the partition serve as the building-blocks for the construction of the cost function $F_{[N]}$. For every cluster $c_{t_s}$ in the partition there are three stabilizers associated to it: $G_{t_s+1}=Z_{t_s}X_{t_s+1}Z_{t_s+2}$, $G_{t_s+2}=Z_{t_s+1}X_{t_s+2}Z_{t_s+3}$ and $G_{t_s+1,t_s+2}=G_{t_s+1}G_{t_s+2}=Z_{t_s}Y_{t_s+1}Y_{t_s+2}Z_{t_s+3}$. To each of them we associate three projectors
\begin{eqnarray}
Q_{t_s}&=&\frac{\openone+G_{t_s+1}}{2},\\
W_{t_s}&=&\frac{\openone+G_{t_s+2}}{2},\\
R_{t_s}&=&\frac{\openone+G_{t_s+1}G_{t_s+2}}{2},
\end{eqnarray}
that project on the $+1$ outcome. We associate the following measurement settings with each projector
\begin{equation}\label{cluster sett}
\{ZXZZ,ZZXZ,ZYYZ\},
\end{equation}
and we assign ``success'' to the cluster measurement only if the outcome $+1$ (for the value of measured stabilizer) occurs. Formally speaking, for every cluster we define the following local cost function
\begin{equation}\label{local cost}
F_s=F_{m}^{i_1i_2i_3i_4}=\frac{1}{2}+\frac{1}{2}\left\{
                           \begin{array}{ll}
                             (-1)^{i_1+i_2+i_3}, & \hbox{$m=ZXZZ$;} \\
                             (-1)^{i_2+i_3+i_4}, & \hbox{$m=ZZXZ$;} \\
                             (-1)^{i_1+i_2+i_3+i_4}, & \hbox{$m=ZYYZ$,}
                           \end{array}
                         \right.
\end{equation}
where $s=1\dots L$. Finally, for a given partition $\{c_{t_1},c_{t_2},\dots,c_{t_L}\}$ the overall cost function is defined in the following way
\begin{equation}\label{overall cost}
F_{[N]}=\left\{
          \begin{array}{ll}
            1, & \hbox{$F_1+\dots+F_L\geq (\frac{2}{3}+\delta)L$;} \\
            0, & \hbox{$F_1+\dots+F_L< (\frac{2}{3}+\delta)L$,}
          \end{array}
        \right.
\end{equation}
where $\delta>0$ is some constant we keep at the moment as a free parameter. In other words, we associate the ``success'' to the run if the number of local successes exceeds a certain threshold of $(\frac{2}{3}+\delta)L$.
We have defined all we need to set-up the detection procedure. Firstly, a particular partition $\{c_{t_1},c_{t_2},\dots,c_{t_L}\}$ is randomly generated from the set $\mathcal{C}_L$ (with probability $1/|\mathcal{C}_L|$). Secondly, for each cluster in the partition we pick randomly (with probability $1/3$) one setting from the set \eqref{cluster sett} and execute the corresponding measurement.
The experimental run reveals the sequence of results $F_1,F_2,\dots,F_L$ from which we evaluate $F_{[N]}$ by using \eqref{overall cost}.

Now, we will show that the probability of success vanishes exponentially fast for all separable states as the number of qubits grows. Firstly, for a fixed partition $\{c_{t_1},c_{t_2},\dots,c_{t_L}\}$ it is clear that product states fail to satisfy $F_s=1$ for all three settings $\{ZXZZ,ZZXZ,ZYYZ\}$, because $XZ,ZX,YY$ are locally incompatible on a second and third cluster qubit. Thus, if the settings are uniformly distributed (with probability of $1/3$), on can easily show that the probability of success for individual clusters
\begin{eqnarray}\label{Local Prob for cls}
P_{\rho_{prod}}[F_s=1]&=&\avg{F_s}=\frac{1}{3}\langle Q_{t_s}+W_{t_s}+R_{t_s}\rangle\leq\frac{2}{3},~~~
\end{eqnarray}
for all product states $\rho_{prod}=\rho_1\otimes\dots\otimes\rho_N$. Furthermore, if the input state is a product state, the local cost functions $F_s$ can be seen as independent binary (``0/1'') random variables with $\avg{F_s}\leq2/3$ for all $s=1\dots L$. The overall probability of success reads
\begin{equation}
P_{\rho_{prod}}[F_{[N]}=1]=P_{\rho_{prod}}\left[F_{1}+\dots+F_L\geq \left(\frac{2}{3}+\delta\right) L\right],
\end{equation}
and we recognize in the last equation the probability that the sum of independent random variables $F_1+\dots+F_L$ exceeds the value of $(\frac{2}{3}+\delta) L$. 
As $\avg{F_s}\leq2/3$ we expect that the sum $F_1+\dots+F_L$ cannot exceed $2/3L$ significantly. Similar to the previous example (of singlet state), the Chernoff bound holds (see Appendix for the proof), i.e.
\begin{equation}\label{LCS sep bound}
P_{\rho_{prod}}[F_{[N]}=1]\leq e^{-D(\frac{2}{3}+\delta||\frac{2}{3})L},
\end{equation}
where $D(x||y)$ is the Kullback--Leibler divergence. Furthermore, if the bound holds for all product states, it also holds for their mixtures, i.e. it holds for all separable states. Thus, as long as $L$ grows with $N$ (for example we can set $L=[N/5]$, where $[.]$ denotes the integer part) the probability of success vanishes exponentially fast, for all $\delta>0$. As before, we do not have to fix $\delta$ in advance. Once the experiment has been performed, we can calculate directly from the experimental data $F_1,F_2,\dots$ how much the sum of results deviates from $2L/3$, i.e. we set $\delta=(F_1+\dots+F_L)/L-2/3$, and consequently calculate the bound on probability of success for separable states by using \eqref{LCS sep bound}.
For the case of cluster state preparation $\ket{LCS}$, each local cost function $F_s=1$ deterministically, thus we get $\delta=1/3$. The bound \eqref{LCS sep bound} reduces to
\begin{equation}\label{LCS sep bound1}
P_{\rho_{sep}}[F_{[N]}=1]\leq \left(\frac{2}{3}\right)^L.
\end{equation}
If the number of qubits is sufficiently large, even a single-copy of LCS suffices to certify presence of entanglement with high probability. For example, if we want to have a detection probability of at least $95\%$ (i.e. we want to be sure that no separable state has a probability of success more than 5\%) in a single-shot experiment, we get the minimal number of clusters $L_{min}=8$. The lowest number of qubits with such support is $N=24$. Nevertheless, in such a case the set of all partitions $\mathcal{C}_L$ reduces to three only, and for the reason explained bellow (see discussion at the end of this section),
one may want to have $|\mathcal{C}_L|$ significantly larger. For example, already $N=25$ has $|\mathcal{C}_L|=25$, for $N=26$ we get $|\mathcal{C}_L|=117$ etc. 

The scheme can be used not only to detect entanglement, it can be also used to certify the presence of LCS. To see this, note that $P_{\rho}[F_{[N]}=1]=\tr\rho\Pi$, where
\begin{eqnarray}
\Pi=\frac{1}{|\mathcal{C}_L|}\sum_{\{c_{t_1},\dots,c_{t_L}\}\in\mathcal{C}_L}\prod_{s=1}^L\frac{1}{3}(Q_{t_s}+W_{t_s}+R_{t_s}).~~~
\end{eqnarray}
Clearly LCS is the eigenstate $\Pi\ket{LCS}=1\ket{LCS}$ for the maximal eigenvalue. Now, the operator $\Pi$ can be expanded in terms of stabilizers $G_{q_1\dots q_N}$ defined by the equation \eqref{stab equation}. If the set $\mathcal{C}_L$ is sufficiently large the expansion will include all $2^N$ stabilizers. Since the LCS is the only state with $G_{q_1\dots q_N}\ket{LCS}=+1\ket{LCS}$ for all stabilizers, we conclude that LCS is a unique eigenstate of $\Pi$ for eigenvalue 1. Thus, the LCS state is the only state with the maximal probability of success, i.e. $P[F_{[N]}=1]=1$.

We shall briefly comment on the type of entanglement certified by the single-copy detection scheme. Firstly, if we are willing to detect multipartite entanglement, it is very important that the set $\mathcal{C}_L$ is large in size. Recall, that the bound \eqref{LCS sep bound} holds for arbitrary partition from the set $\mathcal{C}_L$ for all separable states. Therefore, if the partition $\{c_{t_1},\dots,c_{t_L}\}$ is fixed and known in advance, the bound \eqref{LCS sep bound} still holds. Nevertheless, for such a case, the following $4$-producible state  $\ket{\phi}=\ket{\psi}_1\ket{\psi}_2\dots \ket{\psi}_L$, where $\ket{\psi}_s$ is the common eigenstate for all three projectors $Q_{t_s}$, $W_{t_s}$ and $R_{t_s}$ (for eigenvalue 1), reveals $F_s=1$ deterministically for every cluster. Consequently we have $P[F_{[N]}=1]=1$ for $\ket{\phi}$ being the input state. Such a state contains localized entanglement on individual clusters (blocks of entanglement). To prevent $\ket{\phi}$ maximizing the probability of success, a random choice of partition from a large set $\mathcal{C}_L$ is necessary. For example, already including additional partition $\{c_{t_1+1},\dots,c_{t_L+1}\}$ obtained by shifting one qubit to the right, prevents $\ket{\phi}$ to be the common eigenstate of $Q_{t_s}$, $W_{t_s}$, $R_{t_s}$ and $Q_{t_s+1}$, $W_{t_s+1}$, $R_{t_s+1}$. On the other hand, if we want $F_{[N]}=1$ deterministically for both partitions, entanglement between neighbouring clusters is needed. Thus, if all partitions of large $\mathcal{C}_L$ are included, the only way to have non-trivial probability of success is to input delocalized entanglement.

Finally, let us briefly explain the generalization to the higher dimensional case. Take an example of a 2D cluster state (known to be universal for quantum computation). Here, one can introduce partitions into $4\times4$ qubit clusters with the corresponding stabilizer projectors (in analogy to $Q_{t_s}$, $W_{t_s}$ and $R_{t_s}$ for LCS) and define the local cost functions. In complete analogy to the 1D case, the 2D detection scheme consists of drawing a random partition followed by a random measurement of local projectors on individual clusters. The separable bound similar to \eqref{LCS sep bound} can be derived. On the other hand, if the 2D cluster state has been prepared, the probability of success is 1.

\subsection*{Ground states of local Hamiltonians}
One of the reasons that single-copy entanglement detection works for cluster states can be associated to the robustness of entanglement with respect to local perturbations. For example, if we measure one or even a group of localized qubits in cluster state, entanglement remains present between the rest qubits.
Ground states of local Hamiltonians are believed to share this property (robustness of entanglement)~\cite{Harrow}, therefore we can expect that they are also amendable to single-copy verification. Here we show that indeed this is the case.

Consider a $L$-local Hamiltonian on some graph of $N$ particles $H=\sum_{k=1}^{N}H^{(k)}$, where $H^{(k)}$ acts on at most $L$ subsystems ($L$ is fixed and independent of $N$). For simplicity reason, we assume the number of local terms $H^{(k)}$ to match the number of particles $N$ (this is common for many physical situations, see for example \cite{Eisert10}). One may consider a more general case where the number of local terms grows as a polynomial function of $N$. Nevertheless, the detection scheme shall work the same way.
Let $\ket{\psi_0}$ is the ground state $H\ket{\psi_0}=N\epsilon_0\ket{\psi_0}$, where $E_0=N\epsilon_0$ is the ground-state energy. We are interested in Hamiltonians that exhibit the so-called entanglement gap $g_{E}=\epsilon_s-\epsilon_0>0$~\cite{Dowling04}, where $\epsilon_s=\frac{1}{N}\min_{\rho_{sep}}\tr H\rho_{sep}$ is the minimal achievable energy (per particle) by a separable state. Furthermore, we assume $g_E$ to be finite and non-zero in the thermodynamical limit, i.e. $0<\lim_{N\rightarrow\infty}g_E<+\infty$. To summarize, we are interested in Hamiltonians where the mean energy $\avg{H}$ can serve as the entanglement witness, i.e. $\avg{H}\geq N\epsilon_s$ for all separable states, whereas at least the ground state violates this bound.

We shall develop a general scheme that works for arbitrary local Hamiltonian. For that reason we introduce a set of tomographically complete measurements for each particle. For example, in the case of qubits, a natural choice are the three complementary measurements defined by $X$, $Y$ and $Z$ Pauli operators. Thus, the set of measurement operators $E^{(k)}_{mi}$ forms a complete basis in the space of observables, i.e. any observable $A^{(k)}$ acting on $k$th subsystem can be decomposed as $A^{(k)}=\sum_{mi}a_{mi}E^{(k)}_{mi}$. Here $m=1\dots M$ and $i=1\dots D$, where $M$ is the number of settings and $D$ is the number of outcomes. Furthermore, in order to simplify the notation, we introduce a new variable $x_k=(m_k,i_k)$ which labels a pair of measurement setting and outcome, hence $E_{x_k}^{(k)}$ refers to $E_{m_ki_k}^{(k)}$. Note that $\sum_{x_k}E_{x_k}^{(k)}=M\openone^{(k)}$.

For a given local Hamiltonian $H=\sum_{k=1}^NH^{(k)}$, operator $H^{(k)}$ acts on at most $L$ neighbouring subsystems (neighbours of $k$ including $k$ itself).
It is convenient to introduce the $N\times L$ ``neighbouring'' matrix $n_{k,l}$, where $n_{k,1},\dots n_{k,L}$ is a sequence of integers labeling all the neighbours of $k$th subsystem (including $k$th subsystem itself) on which the local operator $H^{(k)}$ acts. The ``neighbouring'' matrix can be seen as the list of neighbourhoods $(\mathcal{N}(1),\dots,\mathcal{N}(N))$, where $\mathcal{N}(k)$ denotes the set of all neighbours of $k$. For example, the notation $\{n_{3,1},n_{3,2},n_{3,3}\}=\{2,3,4\}$ means that $H^{(3)}$ acts on subsystems $2,3$ and $4$. Because the set of measurement operators is tomographically complete, each $H^{(k)}$ can be decomposed into the sum of products of local measurement operators
\begin{equation}
H^{(k)}=\sum_{x_{1}\dots x_{L}}h^{(k)}_{x_{1}\dots x_{L}}E^{(n_{k,1})}_{x_{1}}\dots E^{(n_{k,L})}_{x_{L}}.
\end{equation}
The operator $H^{(k)}$ can be completely identified with the tensor $h^{(k)}=h^{(k)}_{x_1\dots x_L}$. Similarly, the full Hamiltonian $H$ reads
\begin{equation}
H=\sum_{x_1\dots x_N}H_{x_1\dots x_N}E^{(1)}_{x_1}\dots E^{(N)}_{x_N},
\end{equation}
where we set $M^{N-L}H_{x_1\dots x_N}=\sum_{k=1}^Nh^{(k)}_{x_{n_{k,1}}\dots x_{n_{k,L}}}$ (the factor $M^{N-L}$ comes because of the normalization $\sum_xE_x=M\openone$). Now we can set-up the detection procedure. Firstly, we pick measurement settings for individual subsystems randomly (i.e. with probability $1/M$) and generate the sequence $\{m_1,\dots,m_N\}$. The measurements are executed on local subsystems and the set of outcomes $\{i_1,\dots,i_N\}$ is obtained. Equivalently, we say that the sequence of random variables $\{x_1,\dots,x_N\}$ is generated, where $x_k=(m_k,i_k)$. Now, we shall define the cost function $F_{[N]}$. It is convenient to define $H_{[N]}=M^N H_{x_1\dots x_N}=M^L\sum_{k=1}^Nh^{(k)}$. A straightforward inspection shows $\avg{H_{[N]}}=\tr\rho H=\avg{H}$, thus the classical random variable $H_{[N]}$ can serve to extract the mean value of Hamiltonian $\avg{H}$. Since $\avg{H}\geq N\epsilon_s$ holds for all separable states, it is natural to chose the following cost function
\begin{equation}\label{Ham cost}
F_{[N]}=\left\{
          \begin{array}{ll}
            1, & \hbox{$H_{[N]}\leq N(\epsilon_s-\delta)$;} \\
            0, & \hbox{$H_{[N]}> N(\epsilon_s-\delta)$,}
          \end{array}
        \right.
\end{equation}
where $0<\delta<\epsilon_s-\epsilon_0=g_{E}$ is constant we keep at the moment as a free parameter. Since the random variable $H_{[N]}$ completely captures properties of Hamiltonian, we expect $H_{[N]}$ not to precede the separable bound $N\epsilon_s$ significantly in a single-shot experiment (provided that $N$ is large).
Indeed, in the Appendix we show that for all separable states $\rho_{sep}$ the following bound holds
\begin{equation}\label{Ham sep bound}
P_{\rho_{sep}}[F_{[N]}=1]\leq\exp\left[-N\kappa^2\delta^2\right],
\end{equation}
where $\kappa>0$ is constant. Thus, the probability of success vanishes exponentially fast with $N$ for all separable inputs. On the other hand, if the ground state $\ket{\psi_0}$ is prepared, we show in the Appendix that the probability of success reaches 1 in the thermodynamical limit, i.e.
\begin{equation}\label{GS bound}
P_{\psi_0}[F_{[N]}=1]\geq1-\frac{\beta^2}{N(g_{E}-\delta)^2},
\end{equation}
where $\beta>0$ is constant. In other words, if $N$ is sufficiently large, the probability of success is close to 1.

There are several points worth of mentioning here. Firstly, the previous example of cluster states can be incorporated in the present scheme, since cluster states can be seen as unique ground-states of local Hamiltonians~\cite{Nest}. Nevertheless, the detection scheme introduced in the previous section is more resource-efficient for cluster states for the following reasons: a) the bound \eqref{LCS sep bound} is more tight than \eqref{Ham sep bound}, and b) the probability of success evaluates to 1 (for the cluster-state input), in contrast to \eqref{GS bound} which reaches 1 asymptotically.
On the other hand, the ground-state detection method has certain practical advantages. Namely, one of the crucial elements for detection is the use of tomographically complete set of measurements. In principle, they can be substituted by a single informationally complete POVM (ICPOVM)~\cite{Renes04}. More precisely, instead of a set of tomographically complete measurements, a single POVM with the measurement operators $E_i$ forming a complete basis in the space of observables, can be used. Thus, an $N$-partite Hamiltonian can be expressed as $H=\sum_{i_1\dots i_N=1}^Dh_{i_1\dots i_N}E_{i_1}^{(1)}\dots E_{i_N}^{(N)}$, where $i_k=1\dots D$ labels the measurement outcome. The properties of Hamiltonian are fully captured by the classical random variable $H_{[N]}=h_{i_1\dots i_N}$ (function of the measurement outcomes $i_1,\dots,i_N$). Now, there is no random sampling of measurement settings, there is only one measurement (ICPOVM) for each particle. The variable $H_{[N]}$ is calculated from the set of measurement outcomes $\{i_1,\dots,i_N\}$. The cost function is defined as \eqref{Ham cost}, and derivation of bounds \eqref{LCS sep bound} and \eqref{Ham sep bound} is essentially the same as before. Formally, both methods are equivalent. Nevertheless, practical advantage of using ICPOVM compared to random sampling of measurement settings can be significant in certain cases, conditioned on the physical implementation of POVM. For example, if ICPOVM is implemented by using additional degrees of freedom (e.g. for the case of single-photons by combining the path and polarization degree of freedom~\cite{Payne}), than, the same, single measurement setting is applied on every local subsystem. This is very convenient when dealing with large-scale quantum systems, for which full manipulation and addressability of individual particles is demanding to achieve.

\subsection*{Tolerance to noise}
Here we analyze the effects of noise on probabilistic entanglement detection. Consider an $N$-partite target state $\rho_0$ with the probability of success $p_0>0$, i.e. there is a chance of $p_0$ to get success (detect entanglement) in a single experimental run (if the state $\rho_0$ has been prepared). This means that in practice, one needs in average $1/p_0$ copies of $\rho_0$ in order to detect entanglement (if the probability of success is $p_0$ than $1/p_0$ experiential runs are needed in average to get ``success'').  Furthermore, let the separable bound \eqref{sep bound} hold, i.e. the probability of success is exponentially small in $N$ for all separable inputs. Now, consider a mixture $\rho=\lambda\rho_{sep}+(1-\lambda)\rho_0$, where $\rho_{sep}$ is an arbitrary separable state and $0<\lambda<1$ quantifies the amount of noise. For example, in many cases the noise can be modeled via the white noise $\rho_{sep}=\openone/d^N$ (here $d$ is the dimension of local Hilbert space) or the product colored noise $\rho_{sep}=\rho_1\otimes\dots\rho_N$~\cite{Ananth2016,DUTTA16}.
The probability of success for such a state is a mixture of probabilities, i.e. $P_{\rho}=\lambda P_{\rho_{sep}}+(1-\lambda)P_{\rho_0}\approx(1-\lambda)p_0$, as long as $(1-\lambda)p_0$ is significantly larger than $P_{\rho_{sep}}=O(exp[-Nc])$. This means that noise affects detection by suppressing the probability of success by the factor $1-\lambda$, for any type of separable noise (i.e. modeled by a separable state). Therefore, one needs in average $\frac{1}{(1-\lambda)p_0}$ experimental runs in order to verify the presence of entanglement. This a strong resistance to noise, as if $(1-\lambda)p_0$ is not exponentially small in $N$ (for example, we consider $(1-\lambda)p_0>0$ constant and independent of $N$), entanglement can be verified with the constant cost in terms of resources (number of copies). On the other hand, the situation with standard detection methods is very different. Typically, a witness tolerates noise bellow a certain critical point, i.e. $\lambda<\lambda_c$. Thus, if noise passes the threshold, the scheme does not work even if an infinite number of resources is available.

To illustrate our findings, let us examine the example of a linear cluster state mixed with the white noise $\rho_{LCS}=\lambda\openone/2^N+(1-\lambda)\ket{LCS}\bra{LCS}$, where $\ket{LCS}$ is the linear cluster state defined by the equation \eqref{stab equation}. The presence of entanglement can be detected via the following set of witness operators~\cite{TG05}
\begin{equation}\label{LCS witness}
W_k=\openone-G_k-G_{k+1},
\end{equation}
where the generators $G_k$ are defined by the equation \eqref{stab equation}. One can easily show that $\avg{W_k}_{sep}\geq0$ for all separable states. In contrast, for the linear cluster state preparation we have $\avg{W_k}_{LCS}=-1$, therefore the witness detects entanglement for $\lambda\leq1/2$~\cite{TG05}. On the other and, if our scheme is applied (see section ``Example of cluster states''), the separable bound is given by the equation \eqref{LCS sep bound}, where $\delta>0$ is a free parameter. As before, we set $\delta=1/3$ and we get the probability of seccess $P_{sep}\leq(\frac{2}{3})^L$ (see equation \eqref{LCS sep bound1}), where $L$ is the size of partitions. For $N$ (and consequently $L$) being sufficiently large, $P_{sep}\approx 0$ is negligible. On the contrary, if the state $\rho_{LCS}$ is prepared, the probability of success is lower bounded by $P_{\rho}=\lambda P_{\openone/2^N}+(1-\lambda)P_{\ket{LCS}}\geq1-\lambda$, where we used $P_{\ket{LCS}}=1$. This means that $1/(1-\lambda)$ copies are sufficient in average in order to get success. For example, if we set $\lambda=1/3$, we need three copies (in average) to detect entanglement, whereas in such a case, the witness \eqref{LCS witness} will fail to detect entanglement even if an infinite number of copies is supplied.

\section*{DISCUSSION}
We introduced a probabilistic technique for resource-efficient entanglement detection in large-scale multiparticle quantum systems. We have shown that for variety of quantum states, probability to detect entanglement (as quantified by the probability of success) approaches one exponentially fast with the size of system, implying that even a single copy suffices to verify entanglement with high probability. Our method promises a dramatic reduction of the resource needed for reliable entanglement verification, therefore it has great potential for practical applications in current and near future experiments aiming at generation and manipulation of massive entanglement.

\medskip\medskip
{\bf Acknowledgments.} The authors thank Nata\v sa Dragovi\'c, Dragoljub Go\v canin, \v{C}aslav Brukner, Lee Rozema and Philip Walther for helpful
comments. A. D. acknowledges the support from the Serbian Ministry of Science (Project ON171035).

\bibliography{Single_copy_detection_Ref}
\appendix
\begin{widetext}
\section{Proof of the separable bounds $(9)$ and $(20)$}
As it has been elaborated in the main text, if the input state is a product state $\rho_{prod}=\rho_1\otimes\dots\otimes\rho_N$, the local cost functions $F_s$ can be seen as the independent binary ($0/1$) random variables with $\avg{F_s}=p_s\leq p$. We set $s=1\dots K$. We proceed by the standard method for proving the Chernoff bound, i.e. by applying the Markov's inequality~\cite{Billingsley}
\begin{equation}
P[X\geq X_0]\leq\frac{\avg{X}}{X_0},
\end{equation}
where $X$ is a positive random variable and $X_0>0$. We set $X=F_1+\dots+F_K$ and $X_0=(p+\delta)K=q K$. For any $t>0$ we have
\begin{eqnarray}\nonumber
P_{\rho_{prod}}[F_{[N]}=1]&=&P_{\rho_{prod}}[X\geq X_0]=P_{\rho_{prod}}[e^{t X}\geq e^{t X_0}]\leq\frac{\avg{e^{t X}}}{e^{t X_0}}.
\end{eqnarray}
Thus
\begin{eqnarray}\label{cher bound}
P_{\rho_{prod}}[F_{[N]}=1]&\leq& \prod_{s=1}^K\frac{\avg{e^{t F_s}}}{e^{tq}}=\prod_{s=1}^K\left(\frac{1-p_s+p_se^{t}}{e^{tq}}\right)\leq\left(\frac{1-p+pe^{t}}{e^{tq}}\right)^K,
\end{eqnarray}
where the last inequality follows from $p_s\leq p$, i.e. $1-p_s+p_se^{t}\leq 1-p+pe^{t}$, for all $s=1\dots K$. The function $f(t)=\frac{1-p+pe^{t}}{e^{tq}}$ attains the minimal value for $t_m=\log\frac{(1-p)q}{(1-q)p}$ or equivalently $e^{t_m}=\frac{(1-p)q}{(1-q)p}$. If we substitute $e^{t_m}$ in the right hand side of \eqref{cher bound} we get
\begin{eqnarray}\label{prod bound1}
P_{\rho_{prod}}[F_{[N]}=1]&\leq& e^{-D(q||p)K}=e^{-D(p+\delta||p)K}.
\end{eqnarray}
The bound holds for any product states ${\rho_{prod}}$. For a separable state ${\rho_{sep}}=\sum_k \lambda_k \rho_{prod}^{(k)}$ we have
\begin{eqnarray}\nonumber
P_{\rho_{sep}}[F_{[N]}=1]=\sum_k \lambda_k P_{\rho_{prod}^{(k)}}[F_{[N]}=1]\leq e^{-D(p+\delta||p)K},\\
\end{eqnarray}
which follows directly from \eqref{prod bound1}. The bound $(9)$ is obtained for $K=N$ and $p=\frac{2}{3}$, whereas
$(20)$ follows for $K=L$ and $p=\frac{2}{3}$.

\section{Proof of the separable bound $(26)$ and the entanglement bound $(27)$ }

Firstly, let us analyze the case of a product input state $\rho_{prod}=\rho_1\otimes\dots\otimes\rho_N$. The probability of success reads
\begin{equation}\label{Ps local Ham}
P_{\rho_{prod}}[F_{[N]}=1]=P_{\rho_{prod}}[H_{[N]}\leq N(\epsilon_s-\delta)]=P_{\rho_{prod}}[h^{(1)}+\dots+h^{(N)}\leq K],
\end{equation}
where we recognize the probability that the sum of random variables $h^{(1)}+\dots+h^{(N)}$ precedes certain bound of $K=\frac{N}{M^L}(\epsilon_s-\delta)$ with $0<\delta<\epsilon_s-\epsilon_0=g_E$. Unlike the case of cluster states, the variables $h^{(k)}$ are not independent (for the case of product inputs), therefore the straightforward application of Chernoff bound is not possible. However, as all $h^{(k)}$ depend only on finite number of $L$ ``neighboring'' variables, we expect to obtain the bound similar to $(9)$.
In order to prove $(26)$, we will use the help of the McDiarmid's inequality~\cite{McDir}:

\begin{theorem} Let $x_1,\dots,x_N$ be independent random variables taking values in the set $\mathcal{X}$. Further, let the function $S_{[N]}: \mathcal{X}^N\mapsto \mathbb{R}$ satisfies
\begin{equation}
\left|S_{x_1\dots x_k\dots x_N}-S_{x_1\dots x'_k\dots x_N}\right|\leq \alpha_k
\end{equation}
for all $x_1,\dots,x_N,x'_k\in\mathcal{X}$, than
\begin{equation}\label{McDiar}
P[S_{[N]}-\avg{S_{[N]}}\geq Q]\leq\exp\left[\frac{-2Q^2}{\sum_{k=1}^N\alpha_k^2}\right],
\end{equation}
for all $Q>0$.
\end{theorem}

Firstly, note that for the case of product inputs, the random variables $\{x_1,\dots,x_N\}$ are independent because the probability distribution
\begin{equation}
P_{x_1\dots x_N}=\frac{1}{M^N}\tr\rho_{prod}E^{(1)}_{x_1}\dots E^{(N)}_{x_N}
\end{equation}
is factorizable. We set $S_{[N]}=-(h^{(1)}+\dots+h^{(N)}$). Furthermore, we label $\mathcal{N}(k)=\{n_{k,1},\dots,n_{k,L}\}$ the set of all neighbors of $k$ and we put $|h^{(k)}_{x_1\dots x_L}|\leq h_{\max}$ for all $k$ and all $x_k$. Since we are dealing with the finite-dimensional Hilbert spaces, $h_{\max}$ is always finite and well defined. We apply the condition for the McDiarmid's theorem and we get

\begin{eqnarray}
\left|S_{x_1\dots x_k\dots x_N}-S_{x_1\dots x'_k\dots x_N}\right|&=&\left|-\sum_{l\in\mathcal{N}(k)}h^{(l)}_{x_{n_{l,1}}\dots x_{n_{l,L}}}+h^{(l)}_{x'_{n_{l,1}}\dots x'_{n_{l,L}}}\right|\\
&\leq&\sum_{l\in\mathcal{N}(k)}\left|h^{(l)}_{x_{n_{l,1}}\dots x_{n_{l,L}}}\right|+\left|h^{(l)}_{x'_{n_{l,1}}\dots x'_{n_{l,L}}}\right|\\
&\leq&2Lh_{\max},
\end{eqnarray}

thus $\alpha_k=2Lh_{\max}$. The inequality \eqref{McDiar} reads
\begin{equation}
P_{\rho_{prod}}[S_{[N]}-\avg{S_{[N]}}\geq Q]\leq\exp\left[\frac{-Q^2}{2NL^2h_{\max}^2}\right],
\end{equation}
for all product states $\rho_{prod}$ and all $Q>0$. Now, we shall obtain the bound on probability of success \eqref{Ps local Ham}. We have
\begin{eqnarray}\nonumber
P_{\rho_{prod}}[F_{[N]}=1]&=&P_{\rho_{prod}}[h^{(1)}+\dots+h^{(N)}\leq K]\\\nonumber
&=&P_{\rho_{prod}}[S_{[N]}-\avg{S_{[N]}}\geq -K-\avg{S_{[N]}}]\\\nonumber
&\leq&\exp\left[\frac{-(K+\avg{S_{[N]}})^2}{2NL^2h_{\max}^2}\right]\\\nonumber
&\leq&\exp\left[\frac{-(K-\frac{N}{M^L}\epsilon_s)^2}{2NL^2h_{\max}^2}\right]\\
&=&\exp\left[-N\kappa^2\delta^2\right],
\end{eqnarray}
where $\kappa^2=1/(2M^{2L}L^2h_{\max}^2)$ and $\delta>0$. The second inequality follows from the separable bound $\avg{S_{[N]}}\leq -\frac{N}{M^L}\epsilon_s$.

On the other hand, if the ground state of $H$ is prepared, we show that $(26)$ holds. Recall that $H_{[N]}=M^{L}\sum_{k=1}^Nh^{(k)}$, thus $\avg{H_{[N]}}=M^L\sum_{k=1}^N\avg{h^{(k)}}$. We start by showing that the variance $\mathrm{Var}[H_{[N]}]$ grows linearly with $N$. By definition $\mathrm{Var}[H_{[N]}]=\avg{H_{[N]}^2}-\avg{H_{[N]}}^2$ which we transform into $\mathrm{Var}[H_{[N]}]=\avg{H_{[N]}^2}-\avg{H^2}+\avg{H_{[N]}}^2-\avg{H}^2+\mathrm{Var}[H]$. Because $\avg{H_{[N]}}=\avg{H}$ and $\mathrm{Var}[H]=0$ (the state $\ket{\psi_0}$ is the ground-state of $H$), we get $\mathrm{Var}[H_{[N]}]=\avg{H_{[N]}^2}-\avg{H^2}$.
The expression for the variance reads

\begin{eqnarray}
\mathrm{Var}[H_{[N]}]&=&\avg{H_{[N]}^2}-\avg{H^2}\\
&=&M^{2L}\left\langle\left(\sum_{k=1}^Nh^{(k)}\right)^2\right\rangle-\avg{H^2}\\
&=&\sum_{j,k=1}^NM^{2L}\avg{h^{(j)}h^{(k)}}-\avg{H^{(j)}H^{(k)}}\\
&=&\sum_{j,k\in\ast}^{}M^{2L}\avg{h^{(j)}h^{(k)}}-\avg{H^{(j)}H^{(k)}},
\end{eqnarray}
where $\ast$ refers to the set of ``crossing terms'' only, i.e. those pairs $(j,k)$ that satisfy $j\in\mathcal{N}(k)$ or $k\in\mathcal{N}(j)$ ($j$ is in the ``neighborhood'' of $k$ or $k$ is in the ``neighborhood'' of $j$). For ``non-crossing terms'', we have $M^{2L}\avg{h^{(j)}h^{(k)}}=M^{2L}\avg{h^{(j)}}\avg{h^{(k)}}=\avg{H^{(j)}}\avg{H^{(k)}}=\avg{H^{(j)}H^{(k)}}$, thus the sum vanishes. Note that the total number of ``crossing terms'' is at most $2NL$, i.e. $\sum_{j,k\in\ast}^{}1\leq2NL$. We can bound particular terms in the sum as
\begin{eqnarray}
|\avg{h^{(j)}h^{(k)}}|&=&|\frac{1}{M^{N}}\bra{\psi_0}\sum_{x_1\dots x_N} h^{(j)}_{x_{n_{j,1}},...,x_{n_{j,L}}} h^{(k)}_{x_{n_{k,1}},...,x_{n_{k,L}}}E_{x_1}^{(1)}\dots E_{x_N}^{(N)}\ket{\psi_0}|\\
&\leq&\frac{1}{M^{N}}\sum_{x_1\dots x_N} |h^{(j)}_{x_{n_{j,1}},...,x_{n_{j,L}}}| |h^{(k)}_{x_{n_{k,1}},...,x_{n_{k,L}}}|\bra{\psi_0}E_{x_1}^{(1)}\dots E_{x_N}^{(N)}\ket{\psi_0}\\
&\leq&\frac{A^2}{M^N}\sum_{x_1\dots x_N}\bra{\psi_0}E_{x_1}^{(1)}\dots E_{x_N}^{(N)}\ket{\psi_0}=A^2,
\end{eqnarray}
where $A=\max_{k,x_s}|h^{(k)}_{x_{n_{k,1}},...,x_{n_{k,L}}}|$. Here we used $\sum_{x_k}E_{x_k}^{(k)}=\sum_{m_k,i_k}E_{m_k,i_k}^{(k)}=M\openone^{(k)}$. Furthermore, if we set $B=\max_{k}|\bra{\psi_0}H^{(k)}\ket{\psi_{0}}|$, we get $|\avg{H^{(j)}H^{(k)}}|\leq B^2$, by the Cauchy-Schwarz inequality. Finally, we apply the inequality $|a-b|\leq|a|+|b|$ and by using the expression for variance given above we get
\begin{eqnarray}\label{variance bound}
\mathrm{Var}[H_{[N]}]&=&\left|\sum_{j,k\in\ast}^{}M^{2L}\avg{h^{(j)}h^{(k)}}-\avg{H^{(j)}H^{(k)}}\right|\\
&\leq&\sum_{j,k\in\ast}^{}M^{2L}|\avg{h^{(j)}h^{(k)}}|+|\avg{H^{(j)}H^{(k)}}|\\
&\leq&\sum_{j,k\in\ast}^{}M^{2L}A^2+B^2\\
&\leq&2NL(M^{2L}A^2+B^2)=\beta^2 N,\\
\end{eqnarray}
with $\beta^2=2L(M^{2L}A^2+B^2)$.

Finally, the probability of success reads
\begin{eqnarray}
P_{\psi_0}[F_{[N]}=1]&=&P_{{\psi_0}}[H_{[N]}\leq M^LK]\\
&=&P_{{\psi_0}}[H_{[N]}-\avg{H_{[N]}}\leq N(\epsilon_s-\epsilon_0-\delta)]\\
&=&P_{{\psi_0}}[H_{[N]}-\avg{H_{[N]}}\leq N(g_E-\delta)]\\
&\geq&P_{{\psi_0}}[H_{[N]}-\avg{H_{[N]}}< N(g_E-\delta)]\\
&=&1-P_{{\psi_0}}[H_{[N]}-\avg{H_{[N]}}\geq N(g_E-\delta)]\\
&\geq&1-P_{{\psi_0}}[H_{[N]}-\avg{H_{[N]}}\geq N(g_E-\delta)]-P_{{\psi_0}}[H_{[N]}-\avg{H_{[N]}}\leq -N(g_E-\delta)]\\
&=&1-P_{{\psi_0}}[|H_{[N]}-\avg{H_{[N]}}|\geq N(g_E-\delta)]\\
&\geq&1-\frac{\mathrm{Var}[H_{[N]}]}{N^2(g_E-\delta)^2}\\
&\geq&1-\frac{\beta^2}{N(g_E-\delta)^2}.
\end{eqnarray}
The second last inequality follows from the Chebyshev's inequality~\cite{Billingsley}.

\section{Example of the set of regular partitions for $L=2$ and $N=6,7,8$ }
Here we list the set of all regular partitions (see main text) for the case $L=2$ and $N=6,7,8$, with $|\mathcal{C}_2|=2,4,12$, respectively:
\begin{eqnarray}
N=6:~~~\mathcal{C}_2=&&\{\{1,2,3,4\}, \{4,5,6,1\}\}, \{\{2,3,4,5\} ,\{5,6,1,2\}\},\\
N=7:~~~\mathcal{C}_2=&&\{\{1,2,3,4\}, \{4,5,6,7\}\}, \{\{2,3,4,5\} ,\{5,6,7,1\}\},\{\{3,4,5,6\} ,\{6,7,1,2\}\},\\ \nonumber
&&\{\{4,5,6,7\} ,\{7,1,2,3\}\},\\
N=8:~~~\mathcal{C}_2=&&\{\{1,2,3,4\}, \{5,6,7,8\}\}, \{\{2,3,4,5\} ,\{6,7,8,1\}\},\{\{3,4,5,6\} ,\{7,8,1,2\}\},\\ \nonumber
&&\{\{4,5,6,7\} ,\{8,1,2,3\}\},\{\{1,2,3,4\}, \{4,5,6,7\}\}, \{\{2,3,4,5\}, \{5,6,7,8\}\},\\ \nonumber
 &&\{\{3,4,5,6\}, \{6,7,8,1\}\},\{\{4,5,6,7\}, \{7,8,1,2\}\}, \{\{ 5,6,7,8\}, \{8,1,2,3\}\},\\ \nonumber
 &&\{\{6,7,8,1\}, \{1,2,3,4\}\}, \{\{7,8,1,2\}, \{2,3,4,5\}\}, \{\{8,1,2,3\}, \{3,4,5,6\}\}.
\end{eqnarray}

\end{widetext}

\end{document}